\documentclass{article}

\usepackage[utf8]{inputenc}
\usepackage[numbers]{natbib}
\usepackage[margin=1in]{geometry}
\usepackage{float}
\usepackage{graphicx}
\usepackage{todonotes}
\usepackage{xspace}
\usepackage{amsmath,amssymb,graphicx,amsthm,dsfont,mathtools}

\usepackage[colorlinks]{hyperref}
\hypersetup{
  linkcolor=blue!70!black,        %
  citecolor=red!70!black,         %
}

\usepackage{cleveref}
\usepackage{paralist}
\usepackage[inline]{enumitem}

\newtheorem{corollary}{Corollary}

\newtheorem{claim}{Claim}

\newtheorem{theorem}{Theorem}
\crefname{table}{Table}{Tables}
\crefname{figure}{Figure}{Figures}
\crefname{theorem}{Theorem}{Theorems}
\crefname{corollary}{Corollary}{Corollaries}
\crefname{observation}{Observation}{Observations}
\crefname{lemma}{Lemma}{Lemmas}
\crefname{example}{Example}{Examples}
\crefname{reduction}{Reduction}{Reductions}
\crefname{construction}{Construction}{Constructions}
\crefname{subsection}{Subsection}{Subsections}
\crefname{section}{Section}{Sections}
\crefname{claim}{Claim}{Claims}
\crefname{proposition}{Proposition}{Propositions}

\theoremstyle{definition}
\newtheorem{definition}{Definition}
\crefname{definition}{Definition}{Definitions}

\usepackage{comment}

\newcommand{\myemph}[1]{{\color{green!30!black}\emph{#1}}}

\newcommand{\agents}{V}
\newcommand{\agentsU}{U}
\newcommand{\agentsW}{W}
\newcommand{\ppp}{\mathcal{P}}
\DeclareMathOperator{\rank}{rank}
\newcommand{\MM}{M}

\newcommand{\decprob}[3]{
  \begin{center}%
    \begin{minipage}{0.96\linewidth}%
      \textsc{#1}\\[0.2ex]
      \textbf{Input:} #2\\[0.2ex]
      \textbf{Question:} #3
    \end{minipage}%
  \end{center}
}

\newcommand{\taskprob}[3]{
  \begin{center}%
    \begin{minipage}{0.96\linewidth}%
      \textsc{#1}\\[0.2ex]
      \textbf{Input:} #2\\[0.2ex]
      \textbf{Task:} #3
    \end{minipage}%
  \end{center}
}

\newcommand{\SR}{\textsc{SR}\xspace}
\newcommand{\SM}{\textsc{SM}\xspace}
\newcommand{\SMI}{\textsc{SMI}\xspace}
\newcommand{\SMTI}{\textsc{SMTI}\xspace}
\newcommand{\SRI}{\textsc{SRI}\xspace}
\newcommand{\SRTI}{\textsc{SRTI}\xspace}
\newcommand{\maxSMTI}{\textsc{Max}-\SMTI}
\newcommand{\maxSMTIlong}{\textsc{Maximum Size}-\SMTI}
\newcommand{\minBP}{\textsc{Min-BP}}
\newcommand{\minBPSMI}{\minBP-\SMI}
\newcommand{\minBPSRI}{\minBP-\SRI}
\newcommand{\minBPSMIlong}{\textsc{Minimum Blocking Pairs SMI}}
\newcommand{\minBPSRIlong}{\textsc{Minimum Blocking Pairs SRI}}
\newcommand{\minBPSMIdec}{\minBP-\SMI-\textsc{dec}}
\newcommand{\minBPSRIdec}{\minBP-\SRI-\textsc{dec}}

\newcommand{\maxSMTIdec}{\textsc{Max}-\SMTI-\textsc{dec}}

\newcommand{\Clique}{\textsc{Clique}\xspace}

\DeclareMathOperator{\mcost}{cost}
\newcommand{\cost}{\ensuremath{\mcost}}
\DeclareMathOperator{\goal}{goal}
\DeclareMathOperator{\sol}{sol}

\DeclareMathOperator{\OPT}{OPT}

\DeclareMathOperator*{\argmin}{arg\,min}

\newcommand{\Msize}{k}
\newcommand{\Bnr}{\beta}
\newcommand{\kIS}{h}

\newcommand{\tied}{\tau}

\newcommand{\smicounts}{(\kIS + \binom{\kIS}{2}) \cdot F(\kIS + \binom{\kIS}{2} ) + 1}
\newcommand{\smicount}{\ensuremath{[\smicounts]}}

\newcommand{\np}{\ensuremath{\mathsf{NP}}}
\newcommand{\ptas}{\ensuremath{\mathsf{PTAS}}}

\newcommand{\wone}{\ensuremath{\mathsf{W}[1]}}

\newcommand{\won}{\ensuremath{\mathsf{W}}}

\newcommand{\pp}{\ensuremath{\mathsf{P}}}
\newcommand{\xp}{\ensuremath{\mathsf{XP}}}
\newcommand{\fpt}{\ensuremath{\mathsf{FPT}}}
\newcommand{\AS}{\ensuremath{\mathsf{AS}}}

\newcommand{\Ind}{z}
\newcommand{\Indp}{r}
\newcommand{\Indq}{r'}

\newcommand{\seq}[1]{\ensuremath{\langle#1\rangle}}

\newcommand{\vertexselectorU}{\ensuremath{s}}
\newcommand{\vertexselectorW}{\ensuremath{t}}

\newcommand{\edgeselectorU}{\ensuremath{\hat{s}}}
\newcommand{\edgeselectorW}{\ensuremath{\hat{t}}}

\newcommand{\vertexagentU}{\ensuremath{u}}
\newcommand{\vertexagentW}{\ensuremath{w}}

\newcommand{\xU}{\ensuremath{x}}
\newcommand{\yW}{\ensuremath{y}}

\newcommand{\edgeagentU}{\ensuremath{\pi}}
\newcommand{\edgeagentUU}{\ensuremath{\Pi}}
\newcommand{\edgeagentW}{\ensuremath{\theta}}
\newcommand{\edgeagentWW}{\ensuremath{\Theta}}

\newcommand{\enn}{\ensuremath{\hat{n}}}
\newcommand{\emm}{\ensuremath{\hat{m}}}

\tikzstyle{edge} = [gray!60!black]
\tikzstyle{label} = [inner sep=1.5pt, fill=white]

\tikzstyle{agent} = [draw, circle, fill=black, minimum size=1.4ex, inner sep=1pt, text centered, align=center]
\tikzstyle{agentU} = [draw=red!60!black, circle, fill=red!60!black, minimum size=1.2ex, inner sep=1pt, text centered, align=center]
\tikzstyle{agentW} = [draw=blue!60!black, rectangle, fill=blue!60!black, minimum size=1ex, inner sep=1pt, text centered, align=center]

\colorlet{bigweightc}{red!80!black}
\colorlet{bestc}{green!80!black}

\colorlet{firstcol}{green!50!black}
\colorlet{secondcol}{blue!90!black}
\colorlet{thirdcol}{red!80!black} 

\usepackage{authblk}

\author[1]{Jiehua Chen}
\author[2]{Sanjukta Roy}
\author[1]{Sofia Simola}

\affil[1]{TU Wien, Austria}
\affil[2]{Indian Statistical Institute, India}

\title{\fpt-Approximability of Stable Matching Problems}
\date{}

\begin{document}

\maketitle

\begin{abstract}
  We study parameterized approximability of three optimization problems related to stable matching:
  \begin{inparaenum}[(1)]
    \item \minBPSMI: Given a stable marriage instance and a number~$\Msize$, find a size-at-least-$k$ matching that minimizes the number~$\Bnr$ of blocking pairs;
    \item \minBPSRI: Given a stable roommates instance, find a matching that minimizes the number~$\Bnr$ of blocking pairs;
    \item \maxSMTI: Given a stable marriage instance with preferences containing ties, find a maximum-size stable matching.
  \end{inparaenum}

  The first two problems are known to be NP-hard to approximate to any constant factor and \won[1]-hard with respect to~$\Bnr$, making the existence of an EPTAS or \fpt-algorithms unlikely.
  We show that they are \wone-hard with respect to $\Bnr$ to approximate to any function of~$\Bnr$.
  This means that unless \fpt{}={}\wone, there is no \fpt-algorithm with respect to~$\Bnr$ that can decide whether a given instance has a matching with at most $\Bnr$ or at least $F(\Bnr)$ blocking pairs for every function~$F$ depending only on~$\Bnr$.
  The last problem (\maxSMTI) is known to be \np-hard to approximate to factor-$\frac{29}{33}$ and \won[1]-hard with respect to the number of ties. 
  We complement this and present an \fpt-approximation scheme for the parameter~``number of agents with ties''.
\end{abstract}

\section{Introduction}
Stable matchings play a fundamental role in various applications ranging from job markets (e.g., the National Resident Matching Program),
over education (e.g., school choice or college admission),
to health (e.g., kidney exchange programs).
The general idea is to match the agents into pairs so no two agents can form a blocking pair. 
A blocking pair is a pair of two agents that are not matched with each other but prefer each other over the status under the matching. 

In the classical \textsc{Stable Marriage} (\SM) problem introduced by Gale and Shapley~\cite{gale1962college}, we are given two disjoint sets of agents, each having a complete and strict preference list (aka.\ ranking) over the agents of the other set, and we aim to match every agent from one set to an agent of the other set so that no blocking pairs are induced.
Gale and Shapley show that every instance of the \SM\ problem admits a stable matching and provide a linear-time algorithm to find one.

\SM\ has many relevant extensions.
First, the preference lists of the agents may be \emph{incomplete}, allowing them to express unacceptability towards other agents.
This results in the setting of \textsc{Stable Marriage with Incomplete Preferences} (\SMI).
A stable matching  still always exists for each \SMI\ instance, albeit not every agent is matched; the same set of agents are matched in all stable matchings though. 
A key challenge in \SMI\ is the trade-off between the size and stability: Requiring more agents to be matched will induce blocking pairs, destroying stability \cite[Rural Hospital Theorem]{roth1986allocation}.
This motivates the problem of \minBPSMIlong\ which aims to find a matching of size at least a given number~$\Msize$ such that the number of blocking pairs is minimized. 
Prior work has considered both exact and approximation algorithms~\cite{biro2010size,hamada2009improved,guptaASM}, showing that the problem remains computationally challenging. 
For instance, \minBPSMI\ is already \np-hard when the preference lists are of length at most three~\cite{biro2010size}.
Moreover, it is \np-hard to approximate to some constant factor~\cite{biro2010size,hamada2009improved}.

Another extension of \SM is to have just one set of agents, allowing each agent to rank any other agent.
In this case, we obtain the setting of \textsc{Stable Roommates} (\SR).
Unlike in \SM\ and \SMI, stable matchings are not guaranteed to exist.
\citet{Irving1985} designs a linear-time algorithm to decide whether a given instance of \SR\ admits a stable matching. 
To deal with the non-existence of stable matching, \citet{biro2012almost} introduce the problem of finding a matching that minimizes the number of blocking pairs, called \minBPSRI.
Unfortunately, \minBPSRI is \won[1]-hard with respect to the number~$\Bnr$ of blocking pairs~\cite{chen_roommates} and \np-hard to approximate~\cite{biro2012almost}.

A third variant of \SM\ is to allow agents to have both \emph{incomplete} and \emph{indifferent} preferences and between agents.
In this case, a  stable matching\footnote{When preferences have ties, our definition of stability is also known as weak stability, distinguishing it from the case of strong stability or super stability~\cite{Manlove2013}.
  For strong stability, for each pair of agents \emph{at least one} agent in the pair must prefer his partner over the other agent in the pair.
  For super stability,  for each pair of agents \emph{each agent} in the pair must prefer his partner over the other agent in the pair.} still exists, but different stable matchings may have different sizes.  
A commonly studied goal for incomplete preferences with ties is to find a stable matching with maximum size. 
This gives rise to the setting of \textsc{Maximum-Size Stable Marriage with Incomplete preferences and Ties} (\maxSMTI).
\maxSMTI is \np-hard to approximate~\cite{manlove2002hard} and \won[1]-hard with respect to the number of ties~\cite{marx2010parameterized}. 

The above inapproximability and parameterized intractability results beg the study of approximation algorithms combined with fixed-parameter tractable (\fpt) algorithms.
Given the inherent computational hardness of \minBPSMI, \minBPSRI, and \maxSMTI, parameterized complexity provides a natural framework to explore efficient algorithms when certain problem parameters are small.
In this work, we investigate \fpt-approximability for both minimizing the number of blocking pairs and maximizing the size of stable matchings in presence of ties.

\subsection{Our Contributions}
We show that \minBPSMI and \minBPSRI remain resistant to \fpt-approximability.
More precisely, both problems are \won[1]-hard with respect to $\Bnr$ to approximate to a factor $F(\Bnr)\cdot n^{O(1)}$, where $n$ is the number of agents,  $\Bnr$ is the number of blocking pairs, $F$ is an arbitrary computable function.
In other words, it is unlikely to obtain an \fpt\ algorithm with respect to~$\Bnr$ that can distinguish between instances that have at most $\Bnr$ or at least $F(\Bnr)$ blocking pairs.
This hardness result for \minBPSMI holds even when we require the matching to be perfect, i.e., every agent has to be matched.
On the other hand, when preferences can be incomplete and may have ties, we provide a simple \fpt-approximation scheme for \maxSMTI with respect to the number~$\tied$ of agents with ties, i.e., an approximation scheme that given an instance~$I$ of \SMTI, where $\tied$ agents have preferences containing ties, and an approximation ratio~$\varepsilon$, outputs in $f(\varepsilon, \tied)\cdot |I|^{O(1)}$ time a stable matching with size at least~$\varepsilon\cdot \OPT(I)$, where $\OPT(I)$ denotes the size of a largest stable matching in $I$ and $f$ is a computable function depending only on~$\varepsilon$ and $\tied$.

\subsection{Related Work}
\minBPSMI and \minBPSRI are particularly difficult to approximate.
\citet{biro2010size} and \citet{hamada2009improved} show that \minBPSMI is \np-hard to approximate to factor $n^{1 -\varepsilon}$ for any~$\varepsilon > 0$, where~$n$ is the number of agents.
The result holds even when the desired matching size is $\Msize=\frac{n}{2}$, i.e., every agent has to be matched. 
\citet{hamada2011hospitals} extend the result to the \textsc{Hospital-Residents} problem.
\citet{biro2010size} show that \minBPSMI remains \np-hard to approximate to some constant factor~$\delta > 1$, even when the preference lists are of length at most three.
On the positive side, when one side of the preference lists have length at most two, \minBPSMI\ becomes polynomial-time solvable. 
For the roommates setting, \citet{biro2012almost} show that \minBPSRI is \np-hard to approximate to factor $n^{\frac{1}{2} -\varepsilon}$ for any~$\varepsilon > 0$.
They also consider the case when the preference lists have length at most~$d$.
They obtain a dichotomy result: The problem remains \np-hard to approximate to some constant factor even when $d = 3$, but can be approximated within factor~$2d-3$ in general.
If $d\le 2$, then the problem becomes polynomial-time solvable. 

The parameterized complexity of the canonical decision variants of \minBPSMI and \minBPSRI has attracted some study; in what follows we use \minBPSMI and \minBPSRI\ to refer to the decision variants for the sake of brevity. 
\citet{guptaASM} show that \minBPSMI remains \won[1]-hard for the combined parameter $\Bnr + t + d$, where~$\Bnr$ is the number of blocking pairs, $t$ the increase in size to a stable matching of the instance, and $d$ the maximum length of a preference list.
Towards tractability, they present an algorithm for a local search variant of the problem.
More precisely, they provide an \fpt-algorithm for finding a matching that differs from a given matching $\mu$ by at most $q$ edges, has at most $k$ blocking pairs, and has size at least $|\mu|+t$, parameterized by $q+k+t$.
Moreover, they show that this local search problem is \wone-hard parameterized by $k+t$ even when the maximum length of a preference list is three.
\citet{chen_roommates} study the parameterized complexity of two optimization variants of \SRI and \SRTI (Stable Roommates with Ties and Incomplete lists), one being \minBPSRI\ and the other \textsc{Egalitarian SR}, which aims at a stable matching with minimum egalitarian cost (i.e., sum of the ranks of the agents' partners).
They show that \minBPSRIdec\ is \won[1]-hard with respect to~$\Bnr$, even when the preference lists are of length at most five.
\citet{abraham2005almost} show that \minBPSRI\ is \xp\ with respect to~$\Bnr$, i.e., it is solvable in polynomial time for every fixed constant~$\Bnr$.

Now we turn to the last problem~\maxSMTI.
\citet{manlove2002hard} show that \maxSMTI is \np-hard.
\citet{irving2009stable} strengthen this by showing that the hardness remains even when one side has preference lists of length at most three.
From the approximation point of view, \citet{KiralY2011} and \citet{mcdermid20093} show that \maxSMTI can be approximated to factor~$\frac{2}{3}$. 
\citet{yanagisawa2007approximation} show the problem does not admit a factor-$\frac{29}{33}$ approximation (polynomial-time) algorithm unless \pp$=$\np.
\citet{dudycz2021tight} and \citet{huang2015tight} show that under the assumption of one of two different variants of Unique Games Conjecture,
the factor-$\frac{2}{3}$ approximation algorithm by \citet{KiralY2011} and \citet{mcdermid20093} are that best achievable ones.
Further approximation results for \maxSMTI under various restrictions have been presented, see for example~\cite{huang2015tight,csaji2024simple,koenemann2020approximating}.

\citet{marx2010parameterized,adil2018parameterized} investigate parameterized complexity of \maxSMTI.
\citet{marx2010parameterized} show that \maxSMTI is \won[1]-hard with respect to the number of ties, but \fpt\ with respect to the total sum of the lengths of the ties.
They also study fixed-parameter tractability of other variants of \SMTI, such as egalitarian and minimum regret \SMTI.
\citet{adil2018parameterized} show that \maxSMTI admits a size-$O(\Msize^2)$ problem kernel and can be solved in $\Msize^{O(\Msize)}+n^{O(1)}$~time.

Parameterized approximation has also been studied in the context of voting.
\citet{bredereck2026shiftbribery} show that \textsc{Minimum Cost Borda Shift Bribery} admits an \fpt-approximation scheme for the parameter number of shifts.
\citet{skowron2017chamberlin} and \citet{gupta2025more} study the fixed-parameter approximability of multiwinner voting rules.

%

%
%
%
%
%
%
%
%

\section{Preliminaries}
Given a non-negative integer~$t$, we use \myemph{$[t]$} to denote the set~$\{1,\ldots,t\}$.

\paragraph{Preference profiles and stable matchings.} Let $\agents$ be a set of $n$ agents with $V=\{v_1,\dots,v_n\}$.
Each agent $v_i \in \agents$ has a subset~$\agents_i \subseteq \agents \setminus \{i\}$ that it finds \myemph{acceptable} and has a \myemph{preference list} $\succeq_i$ on $\agents_i$ which is a weak order over $\agents_i$.
We assume the acceptability to be mutualm, i.e., an agent~$x$ is in the preference list of another agent~$y$ if and only~$y$ is also in the preference list of~$x$.
For two acceptable agents~$x, y \in \agents_i$ of~$v_i$, we say that $v_i$ \myemph{weakly prefers} $x$ over $y$ if $x \succeq_i y$.
If not also $y \succeq_i x$, we write $x \succ y$ and say that $v_i$ \myemph{strictly prefers} (or simply \myemph{prefers}) $x$ over $y$.
If $x \succeq_i y$ and $y \succeq_x$, we write $x \sim_i y$, and say that $v_i$ is \myemph{indifferent} between $x$ and $y$, and has \myemph{ties} in his preference list. 

A preference profile~$\ppp = (\agents, (\succeq_i)_{i \in \agents})$ is a collection of the agents and their preference lists.
Associated to each preference profile~$\ppp$ there is an underlying \myemph{acceptability graph}~$G$, which has $\agents$ as its vertex set, and an edge between each pair of agents who find each other acceptable.
We say that $\ppp$ is a \myemph{marriage} instance if the corresponding acceptability graph is bipartite, otherwise it is a \myemph{roommates} instance.
Further, $\ppp$ is \myemph{complete} if the acceptability graph is complete.

We say that $\ppp$ contains \myemph{ties} if there is an agent $v_i \in \agents$ and two other agents $x, y \in \agents_i$ such that $x \sim_i y$. If $\ppp$ has no ties, then we say that it is \myemph{strict}.

For example, assume we have four agents $\{v_1,v_2,v_3,v_4\}$.
Let agent $v_1$'s preference list be $v_2 \succ_1 v_3 \sim_1 v_4$.
This means that agent~$v_1$ prefers $v_2$ to both $v_3$ and $v_4$, but is indifferent between $v_3$ and $v_4$.

The \myemph{rank} of an agent~$x$ in the preference list of another agent~$v_i$ denotes the number of agents that $v_i$ strictly prefers over $x$:
\myemph{$\rank^{\ppp}_i(x) \coloneqq |\{y \in \agents_i \mid y \succ_i x\}|$}.
We omit the superscript $\ppp$ if it is clear from the context.

Given a preference profile~$\ppp = (\agents, (\succeq_i)_{i \in \agents})$, a \myemph{matching} $\MM \subseteq E(G)$ is a subset of disjoint edges in~$G$. %
Given an pair~$\{v_i, v_j\} \in \MM$, we say that $v_j$ is the \myemph{partner} of $v_i$, denoted $\MM(v_i)$, and we say that $v_i$ and $v_j$ are \myemph{matched}.
If there is no pair~$e \in \MM$ such that $v_i \in e$, then we say that $v_i$ is \myemph{unmatched}, denoted as $\MM(v_i)=\bot$.
We assume that every agent~$v_i$ prefers to be matched with an acceptable agent over being unmatched~$\bot$.

Given a matching $\MM$ of $\ppp$,
a pair $\{v_i,v_j\} \in E(G)$ of agents is a \myemph{blocking} pair of and \myemph{blocks} $\MM$ if
$\{v_i,v_j\}\notin \MM$ and 
the following two conditions hold:
\begin{compactenum}[(i)]
  \item $v_i$ is unmatched or prefers $v_j$ to~$\MM(v_i)$, and 
  \item $v_j$ is unmatched or prefers~$v_i$ to~$\MM(v_j)$.
\end{compactenum}
We call a matching $\MM$ \myemph{stable} if there is no blocking pair; otherwise we call it \myemph{unstable}. If every agent is matched, then we call the matching \myemph{perfect}.

\paragraph{Problem definitions.} 
We are ready to formally define the computational problems studied in this paper.

\taskprob{\minBPSMIlong~({\normalfont\text{in short}} \minBPSMI)}
{A marriage instance~$\ppp = (\agentsU \cup \agentsW, (\succeq_i)_{i \in \agentsU \cup \agentsW})$
  with strict preferences and a non-negative integer~$\Msize$.} %
{Find a matching $\MM$ of size~$|\MM| \geq \Msize$ with minimum number of blocking pairs.}

\taskprob{\minBPSRIlong~({\normalfont\text{in short}} \minBPSRI)}
{A roommates instance~$\ppp = (\agents, (\succeq_i)_{i \in \agents})$ with strict preferences.} %
{Find a matching with minimum number of blocking pairs.}

The standard parameterized decision variants of \minBPSMI\ and \minBPSRI, called \minBPSMIdec\ and \minBPSRIdec, additionally have as input a number~$\Bnr$, and asks whether
there exists a solution with at most~$\Bnr$ blocking pairs.

Now, we turn to preferences with ties.
\taskprob{\maxSMTIlong~(in short\maxSMTI)}
{A marriage instance~$\ppp = (\agentsU \cup \agentsW, (\succeq_i)_{i \in \agentsU \cup \agentsW})$
  with incomplete preferences possibly containing ties.}
{Find a stable matching $\MM$ with maximum size.}

The parameterized decision variant of~\maxSMTI, called \maxSMTIdec, has an additional parameter~$\Msize$ in the input and asks whether there exists a stable matching of size at least~$\Msize$. 

For more details on algorithmic complexity  of stable matching problems, we refer to the textbook by \citet{Manlove2013}.

\paragraph{Fixed-parameter approximability.}
We assume basic knowledge of parameterized complexity and refer the textbook by~\citet{CyFoKoLoMaPiPiSa2015} for more details.

The definitions for fixed-parameter approximability are based on the definitions by~\citet{marx2008parameterized}.

We define an \myemph{optimization problem} as a 4-tuple $(\Pi, \sol, \mcost, \goal)$, where
\begin{compactitem}[--]
  \item $\Pi$ is a set of instances,
  \item $\sol$ is a function that assigns to each instance~$I$ a set consisting of all feasible solutions, i.e., for each~$I \in \Pi$, $\sol(I)$ is a collection of feasible solutions,
  \item $\cost$ is a function that assigns to each instance~$I$ and each feasible solution~$x\in \sol(I)$ an integer that describes the quality of the solution,
  \item $\goal$ is $\min$ for minimization problems and $\max$ for maximization problems.
\end{compactitem}

For instance, for the \minBPSMI\ problem, $\Pi$ consists of all tuples~$I=(\ppp, \Msize)$ such that $\ppp$ is a marriage instance with strict preferences, $\Msize$ is a non-negative integer,
$\sol(I)$ consists of all matchings of size at least~$\Msize$,
$\cost(\MM, I)$ is the number of blocking pairs in~$\MM$,
and $\goal=\min$.

\begin{definition}[Polynomial-time approximation algorithms]
  To define approximation algorithms, let us first consider minimization problems.
  Let $(\Pi, \sol, \mcost, \goal)$ be a minimization problem and $I$ an instance of $\Pi$.
  
  For a given constant~$\rho \ge 1$, we say that an algorithm~$A$ is a \myemph{factor-$\rho$ approximation algorithm} for~$(\Pi, \sol, \mcost, \goal)$ if it outputs a solution~$x$ with $\mcost(x, I) \leq \rho \cdot \OPT$, where $\OPT(I)$ is the cost of an optimal solution of $I$, i.e., $\displaystyle \OPT(I)=\argmin_{x \in \sol}\cost(x, I)$.
  If $(\Pi, \sol, \mcost, \goal)$ is a maximization problem instead,
  then we require $\mcost(x, I) \geq \rho\cdot \OPT(I)$ to holds for a given~$\rho \le 1$. %
  We say that $(\Pi, \sol, \mcost, \goal)$ is \myemph{polynomial-time approximable} to a factor of~$\rho$ if it admits a factor-$\rho$ approximation algorithm that runs in polynomial time.
  
 We say that $(\Pi, \sol, \mcost, \goal)$ admits a \myemph{polynomial-time approximation scheme} (\ptas) if for every approximation error~$\varepsilon > 0$, it is polynomial-time approximable to a factor of~$1+\varepsilon$ (for minimization) and $1-\varepsilon$ (for maximization).
\end{definition}

\begin{definition}[Fixed-parameter approximation algorithms]
  Let $(\Pi, \sol, \mcost, \goal)$ be an optimization problem and $I$ an instance of $\Pi$.
  We say that $(\Pi, \sol, \mcost, \goal)$ is \myemph{\fpt-approximable to factor~$\rho$} with respect to a parameter~$k$ if it admits a factor-$\rho$ approximation algorithm that runs in time $f(\rho, k)\cdot n^{O(1)}$, where $f$ is a computable function on~$k$ and $\rho$.

  Consequently, we say that $(\Pi, \sol, \mcost, \goal)$ admits an \myemph{\fpt-approximation scheme} (\fpt-AS) with respect to a parameter~$k$
  if for every $\varepsilon > 0$ it is \fpt-approximable to factor~$(1+\varepsilon)$ (for minimization) and $(1-\varepsilon)$ (for maximization) with respect to~$k$.
\end{definition}

To show that no \fpt-approximation scheme exists, we need a specific reduction, called 
\myemph{gap-producing parameterized reduction}~\cite{feldmann2020survey-param-approx}.
The idea is to provide a parameterized reduction to a gap version of an optimization problem.
Below, we only define the reduction for minimization problems.
The one for maximization problems is analogous.
\begin{definition}[gap-producing parameterized reduction]
  Let $X=(\Pi, \sol, \mcost, \goal)$ be a minimization problem and let $k$ be the natural parameter of~$X$ to turn it into a decision problem.
  Let $\alpha\colon \mathds{N} \to \mathds{N}$ be a computable function.

  A reduction~$R$ from a parameterized problem~$\hat{\Pi}$ with parameter~$\kIS$ to our optimization problem~$X$ parameterized by~$k$ is called an \myemph{$\alpha$-gap producing parameterized reduction} if
  for every instance~$\hat{I}$ of~$\hat{\Pi}$ with parameter~$\kIS$,
  $R$ produces in $f(\kIS)\cdot |\hat{I}|^{O(1)}$ time an instance~$I$ of~$\Pi$ with parameter~$k$ such that
  \begin{compactenum}[(i)]
    \item $k$ is bounded by a function of~$\kIS$,
    \item if $(\hat{I}, \kIS)$ is a yes-instance, then $I$ admits a solution~$x$ with $\mcost(x,I) \le k$;
    \item if $(\hat{I}, \kIS)$ is a no-instance,
    then every solution~$x$ of~$I$ has~$\mcost(x,I) > \alpha(k)\cdot k$.
  \end{compactenum}
\end{definition}

The existence of an $\alpha$-gap-producing parameterized reduction from a parameterized intractable (e.g., \wone-hard) problem shows that a factor-$\alpha$ \fpt\ approximation algorithm is unlikely, unless $\fpt=\wone$.

\section{\fpt\ Inapproximability of \minBPSMI and \minBPSRI}
In this section, we show that \minBPSMI and \minBPSRI remain parameterized inapproximable with respect to~$\Bnr$, where $\Bnr$ denotes the number of blocking pairs in the natural parameterization variant of the problems.
The idea is to present a gap producing reduction from the \Clique problem that is \won[1]-hard with respect to the size~$\kIS$ of clique.

\decprob{\Clique}
{A directed graph $G = (V, E)$ and a non-negative number~$\kIS$.}
{Does $G$ admit a \myemph{clique}~$K\subseteq V$ of size at least~$\kIS$, , i.e., $G[K]$ is a complete subgraph (aka.\ \myemph{clique}) of size at least~$\kIS$?}

We remark that the reduction presented here is a modified version of Theorem 1 by \citet{guptaASM}.
In their reduction, the vertex gadget is connected to the edge gadget by an edge.
However, we add $\smicount$ cycles of length four to connect a vertex gadget with an ``incident'' edge gadget instead to boost the gap.
\begin{theorem}\label{thm:asm_inapprox}
  Let $F\colon \mathds{N} \to \mathds{N}$ be an arbitrary computable function.
  Given an instance~$(\ppp, \Msize, \Bnr)$ of \minBPSMIdec\, it is \wone-hard to distinguish between the following two cases:
  \begin{compactenum}[(i)]
    \item $\ppp$ admits a size-$\Msize$ matching with at most~$\Bnr$ blocking pairs.
    \item Every size-at-least~$\Msize$ matching of $\ppp$ has at least $F(\Bnr)\cdot \Bnr$ blocking pairs.
  \end{compactenum}
\end{theorem}

\begin{figure}[t]
  \centering
  \begin{tikzpicture}
    \def \xss {24ex}
    \def \xs {6ex}
    \def \ys {6ex}
    \def \yss {8ex}
    
    \node[agentW] (wi) {};
    \node[agentU, left = \xss of wi] (hui) {};
    \node[agentU, above = \ys of wi] (ui) {};
    \node[agentW, above = \ys of hui] (hwi) {};

    \node[agentU, below left = \yss and 1.2*\xs of wi] (s1) {};
    \node[agentU, below right = \yss and 1.2*\xs of wi] (sk) {};

    \node[agentW, below left = \yss and 1.2*\xs of hui] (t1) {};
    \node[agentW, below right = \yss and 1.2*\xs of hui] (tk) {};

    \node[agentW, below right = \yss and 1.2*\xs of s1] (wj) {};
    \node[agentU, below right = \yss and 1.2*\xs of t1] (huj) {};

    \node[agentU, below = \ys of wj] (uj) {};
    \node[agentW, below = \ys of huj] (hwj) {};

    \node[agentW, right = 0.7*\xss of ui] (fxi) {};
    \node[agentU, right = 0.6*\xss of fxi] (eui) {};

    \node[agentW, below = 1.4*\ys of fxi] (hfxi) {};
    \node[agentU, right = 0.6*\xss of hfxi] (heui) {};

    \node[agentW, right = 0.7*\xss of uj] (fxj) {};
    \node[agentU, right = 0.6*\xss of fxj] (euj) {};

    \node[agentW, above = 1.4*\ys of fxj] (hfxj) {};
    \node[agentU, right = 0.6*\xss of hfxj] (heuj) {};

    \path (heuj) edge[draw=none] node[midway, agentW, xshift=0.6*\xss] (ye) {} (heui);
    \node[agentU, above right = \ys and \xs of ye] (xe) {};

    \node[agentU, below right = \ys*1.5 and 0.4*\xs of ye] (hs1) {};
    \node[agentU, below right = \ys*1.5 and 2.5*\xs of ye] (hskk) {};

    \node[agentW, above left = \ys*1.5 and 1.5*\xs of xe] (ht1) {};
    \node[agentW, above right = \ys*1.5 and 1.5*\xs of xe] (htkk) {};

    \path (hs1) edge[draw=none] node[] {$\cdots$} (hskk);
    \path (ht1) edge[draw=none] node[] {$\cdots$} (htkk);
    
    \foreach \l / \p / \d / \n in {
      wi/left/0pt/{$\vertexagentW_i$}, hui/left/0pt/{$\hat{\vertexagentU}_i$},
      ui/above/0pt/{$\vertexagentU_i$}, hwi/above/0pt/{$\hat{\vertexagentW}_i$},
      wj/left/0pt/{$\vertexagentW_j$}, huj/left/0pt/{$\hat{\vertexagentU}_j$},
      uj/below/0pt/{$\vertexagentU_j$}, hwj/below/0pt/{$\hat{\vertexagentW}_j$},
      s1/left/0pt/{$s_1$}, 
      sk/right/0pt/{$s_{\kIS}$}, 
      t1/left/0pt/{$t_1$}, 
      tk/right/0pt/{$t_{\kIS}$},
      fxi/above/0pt/{$\edgeagentW^{\vertexagentU_i}_e[\Indp]$},
      hfxi/left/0pt/{$\hat{\edgeagentW}^{\vertexagentU_i}_e[\Indp]$},
      eui/above/0pt/{${\edgeagentU}^{\vertexagentU_i}_e[\Indp]$},
      heui/below/0pt/{~~~$\hat{\edgeagentU}^{\vertexagentU_i}_e[\Indp]$},
      fxj/below/0pt/{$\edgeagentW^{\vertexagentU_j}_e[\Indp]$},
      hfxj/left/0pt/{$\hat{\edgeagentW}^{\vertexagentU_j}_e[\Indp]$},
      euj/below/0pt/{${\edgeagentU}^{\vertexagentU_j}_e[\Indp]$},
      heuj/above/0pt/{~~~~$\hat{\edgeagentU}^{\vertexagentU_j}_e[\Indp]$},
      ye/right/0pt/{$y_e$},%
      xe/right/0pt/{$x_e$},%
      hs1/below/0pt/{$\hat{s}_1$},%
      hskk/below/0pt/{$\hat{s}_{\binom{\kIS}{2}}$},%
      ht1/above/0pt/{$\hat{t}_1$},%
      htkk/above/0pt/{$\hat{t}_{\binom{\kIS}{2}}$}%
    } {
      \node[\p = \d of \l] {\n}; 
    }
    
    \foreach \s / \t / \ag in {ui/hwi/0, hwi/hui/0, ui/wi/0,uj/hwj/0, hwj/huj/0, uj/wj/0, hui/t1/15,hui/tk/-15, huj/t1/-15,huj/tk/15, wi/s1/15,wi/sk/-15,wj/s1/-15,wj/sk/15,
      ui/fxi/0, uj/fxj/0,
      fxi/eui/-10, fxi/heui/0, hfxi/eui/0, heui/heui/0, hfxi/heui/10,
      fxj/euj/10, fxj/heuj/0, hfxj/euj/0, heuj/heuj/0, hfxj/heuj/-10,
      ye/hs1/0, ye/hskk/0, xe/ht1/0, xe/htkk/0,
      ye/eui/0, ye/euj/0,
      xe/ye/0} {
      \draw (\s) edge[bend right=\ag] (\t);
    }
    \foreach \s / \t / \p / \l / \col / \ag in {
      ui/hwi/0.15/{$0$}/{bestc}/0,
      hwi/ui/0.15/{1}/{black}/0, hwi/hui/0.2/0/{bestc}/0, hui/hwi/0.2/0/{bestc}/0,
      hui/t1/0.3/{$\ge 1$}/{black}/15,
      hui/tk/0.3/{$\ge 1$}/{black}/-15,
      hui/t1/0.8/{$\le \enn$}/{black}/15,
      hui/tk/0.8/{$\le \enn$}/{black}/-15,
      huj/t1/0.3/{$\ge 1$}/{black}/-15,
      huj/tk/0.3/{$\ge 1$}/{black}/15,
      huj/t1/0.8/{$\le \enn$}/{black}/-15,
      huj/tk/0.8/{$\le \enn$}/{black}/15,
      wi/s1/0.3/{$\ge 1$}/{black}/15,
      wi/sk/0.3/{$\ge 1$}/{black}/-15,
      wi/s1/0.8/{$\le \enn$}/{black}/15,
      wi/sk/0.8/{$\le \enn$}/{black}/-15,
      wj/s1/0.3/{$\ge 1$}/{black}/-15,
      wj/sk/0.3/{$\ge 1$}/{black}/15,
      wj/s1/0.8/{$\le \enn$}/{black}/-15,
      wj/sk/0.8/{$\le \enn$}/{black}/15,
      ui/wi/0.2/$> L$/{bigweightc}/0,
      ui/wi/0.2/$> L$/{bigweightc}/0,
      wi/ui/0.2/0/{bestc}/0,
      uj/hwj/0.15/{$0$}/{bestc}/0,
      hwj/uj/0.15/{1}/{black}/0, hwj/huj/0.2/0/{bestc}/0, huj/hwj/0.2/0/{bestc}/0,
      uj/wj/0.2/$> L$/{bigweightc}/0,
      wj/uj/0.2/0/{bestc}/0,
      s1/sk/0.5/{$\cdots$}/{black}/0,%
      t1/tk/0.5/{$\cdots$}/{black}/0,%
      ui/fxi/0.15/{$\le L$}/{black}/0, uj/fxj/0.15/{$\le L$}/{black}/0,
      ui/fxi/0.85/{$1$}/{black}/0, uj/fxj/0.85/{$1$}/{black}/0,
      fxi/eui/0.15/{$2$}/{black}/-10,%
      fxi/eui/0.85/{$0$}/{bestc}/-10,
      fxi/heui/0.15/{$0$}/{bestc}/0, %
      fxi/heui/0.85/{$1$}/{black}/0, %
      hfxi/eui/0.15/{$0$}/{bestc}/0, 
      hfxi/eui/0.85/{$2$}/{black}/0,
      hfxi/heui/0.15/{$2$}/{black}/10, %
      hfxi/heui/0.85/{$0$}/{bestc}/10,%
      fxj/euj/0.15/{$2$}/{black}/10,%
      fxj/euj/0.85/{$0$}/{bestc}/10,
      fxj/heuj/0.15/{$0$}/{bestc}/0, %
      fxj/heuj/0.85/{$1$}/{black}/0, %
      hfxj/euj/0.15/{$0$}/{bestc}/0, 
      hfxj/euj/0.85/{$2$}/{black}/0,
      hfxj/heuj/0.15/{$2$}/{black}/-10, %
      hfxj/heuj/0.85/{$0$}/{bestc}/-10,
      ye/hs1/0.35/{$> L$}/{bigweightc}/0,%
      ye/hskk/0.35/{$> L$}/{bigweightc}/0,
      ye/hs1/0.85/{$\le \emm$}/{black}/10,%
      ye/hskk/0.85/{$\le \emm$}/{black}/-10,
      xe/ht1/0.25/{$\ge\!1$}/{black}/10,%
      xe/htkk/0.25/{$\ge\!1$}/{black}/-10,
      xe/ht1/0.8/{$\le \emm$}/{black}/0,%
      xe/htkk/0.8/{$\le \emm$}/{black}/0,
      ye/eui/0.2/{$\le L$}/{black}/0,
      ye/euj/0.2/{$\le L$}/{black}/0,
      ye/eui/0.85/{$1$}/{black}/0,
      ye/euj/0.85/{$1$}/{black}/0,
      ye/xe/0.2/{$0$}/{bestc}/0,
      ye/xe/0.8/{$0$}/{bestc}/0%
    } {
      \path (\s) edge[draw=none, bend right=\ag] node[pos=\p, fill=white, inner sep=1pt, text=\col] {\scriptsize \l} (\t);
    }
  \end{tikzpicture}
  \caption{Illustration of the main construction for the gap reduction for \cref{thm:asm_inapprox}. Here, let $\Indp \in [\smicounts]$ and $e\in E$ with $e=\{v_i,v_j\}$.}
  \label{fig:thm1}
\end{figure}

\begin{proof}
  We reduce from \Clique, parameterized by the clique size~$\kIS$.
  Let $(G = (V, E), \kIS)$ be an instance of \Clique\ with $V=\{v_1,\dots v_{\enn}\}$ and $E=\{e_1,\dots, e_{\emm}\}$.  
  We construct an instance $(I, \Msize)$ of \minBPSMI as follows; the construction is illustrated in \cref{fig:thm1}.
  For brevity's sake, define $L\coloneqq\smicounts$.
  
  \paragraph{Agents.} We create two disjoint sets of agents, $U$ and $W$.
  \begin{compactitem}[--]
    \item For every ${\Ind} \in [\kIS]$, we create two vertex selector agents~$\vertexselectorU_{\Ind}$ and $\vertexselectorW_{\Ind}$, and add them to $U$ and $W$, respectively.
    \item For every~$p\in [\binom{\kIS}{2}]$,
    we create two edge selector agents~$\edgeselectorU_{p}$ and $\edgeselectorW_{p}$,
    and add them to $U$ and $W$, respectively.

    \item For every vertex~$v_i \in V$, we create four vertex-agents~$\vertexagentU_i$, $\hat{\vertexagentU}_i$, $\vertexagentW_i$, and $\hat{\vertexagentW}_i$,
    and add the first two to~$U$, and the last two to~$W$.
    \item
    For every edge~$e\in E$, we do the following.
    We create two edge-agents~$\xU_e$ and $\yW_e$, and add them to $U$ and~$W$, respectively.
    Additionally, let $v_i$ and $v_j$ denote the two endpoints of~$e$, then,
    we create $8$ groups of edge agents~$\edgeagentU_e^{\vertexagentU_i}[\Indp]$, 1$\edgeagentU_e^{\vertexagentU_j}[\Indp]$, $\hat{\edgeagentU}_e^{\vertexagentU_i}[\Indp]$, $\hat{\edgeagentU}_e^{\vertexagentU_j}[\Indp]$,
    $\edgeagentW_e^{\vertexagentU_i}[\Indp]$, $\edgeagentW_e^{\vertexagentU_j}[\Indp]$, $\hat{\edgeagentW}_e^{\vertexagentU_i}[\Indp]$, $\hat{\edgeagentW}_e^{\vertexagentU_j}[\Indp]$, for $\Indp\in [L]$. 
    We add the first four groups to~$U$ and the last four groups to~$W$.
    Note that these edge agents' preference orders will enforce that matching two adjacent vertices to the selector agents will induce too many blocking pairs (at least~$L$).
  \end{compactitem}
  Summarizing, $U\coloneqq \{\vertexselectorU_{\Ind}\mid \Ind \in [\kIS]\} \cup
  \{\edgeselectorU_{p}\mid p \in [\binom{\kIS}{2}]\} \cup
  \{\vertexagentU_i, \hat{\vertexagentU}_i \mid v_i\in V\} \cup
  \{\xU_e \mid e\in E\}
  \cup \{\edgeagentU^{\vertexagentU_i}_e[\Indp], \edgeagentU^{\vertexagentU_j}_e[\Indp], \hat{\edgeagentU}^{\vertexagentU_i}_e[\Indp],\hat{\edgeagentU}^{\vertexagentU_i}_e[\Indp] \mid e\in E \text{ with } e = \{v_i, v_j\}, \Indp \in [L]\}$
  and $W\coloneqq \{\vertexselectorW_{\Ind} \mid \Ind \in [\kIS]\}
  \cup
  \{\edgeselectorW_{p}\mid p \in [\binom{\kIS}{2}]\} \cup
  \{\vertexagentW_i, \hat{\vertexagentW}_i \mid v_i\in V\} \cup
  \{\yW_e \mid e\in E\}
  \cup
  \{\edgeagentW^{\vertexagentU_i}_e[\Indp], \edgeagentW^{\vertexagentU_j}_e[\Indp], \hat{\edgeagentW}^{\vertexagentU_i}_e[\Indp],\hat{\edgeagentW}^{\vertexagentU_j}_e[\Indp] \mid e\in E \text{ with } e = \{v_i, v_j\}, \Indp \in [L]\}$. 
  In total, we have created $2(\kIS+\binom{\kIS}{2}+ 2\enn+ \emm+ 4\emm\cdot L)$ agents.
  
  \paragraph{Agents' preference orders.} For a given subset~$Q$ of agents, let $\seq{Q}$ denote an arbitrary but fixed linear order of the elements in~$Q$.

  The preference orders of the agents are as follows, where $\Ind \in [\kIS]$, $p\in [\binom{\kIS}{2}]$, $\Indp \in [L]$,  $e\in E$ with $e=\{v_i,v_j\}$, 
  \myemph{$\edgeagentWW(v_i)$} $\coloneqq\{\edgeagentW^{\vertexagentU_{i}}_{e'}[\Indq]\mid v_i \in e' \text{ for some }e' \in E\wedge \Indq \in [L]\}$, and
  \myemph{$\edgeagentUU(e)$} $\coloneqq \{\edgeagentU^{\vertexagentU_i}_e[\Indq], \edgeagentU^{\vertexagentU_j}_e[\Indq] \mid \Indq \in [L]\}$.
  \begin{alignat*}{3}
    \vertexselectorU_{\Ind} &\colon \seq{\{{\vertexagentW}_i \mid v_i\in V\}}, &\qquad \vertexselectorW_{\Ind} &\colon \seq{\{\hat{\vertexagentU}_i \mid v_i\in V\}},\\
    \edgeselectorU_{\Ind} &\colon \seq{\{\yW_{e} \mid e\in E\}}, &\qquad \edgeselectorW_{\Ind} &\colon \seq{\{\xU_e \mid e\in E\}},\\[1ex]
    \vertexagentU_i&\colon \hat{\vertexagentW}_i \succ
    \seq{\edgeagentWW(v_i)} \succ \vertexagentW_i, 
    &
    \vertexagentW_i&\colon \vertexagentU_i \succ  \vertexselectorU_{1}\succ \cdots \succ \vertexselectorU_{\kIS},\\
    \hat{\vertexagentU}_i&\colon \hat{\vertexagentW}_i  \succ \vertexselectorW_{1}\succ \cdots \succ \vertexselectorW_{\kIS},
    &
    \hat{\vertexagentW}_i&\colon \hat{\vertexagentU}_i \succ \vertexagentU_i,\\[1ex]
    \xU_{e} & \colon \yW_e \succ \edgeselectorW_1 \succ \cdots \succ \edgeselectorW_{\binom{\kIS}{2}}, &\qquad
    \yW_e & \colon \xU_e \succ \seq{\edgeagentUU(e)} \succ \edgeselectorU_1 \succ \cdots \succ \edgeselectorU_{\binom{\kIS}{2}}, \\
    \edgeagentU_{e}^{\vertexagentU_i}[\Indp]&\colon \edgeagentW_{e}^{\vertexagentU_i}[\Indp] \succ \yW_e \succ \hat{\edgeagentW}_{e}^{\vertexagentU_i}[\Indp],
    & \edgeagentW_{e}^{\vertexagentU_i}[\Indp] &\colon \hat{\edgeagentU}_{e}^{\vertexagentU_i}[\Indp] \succ \vertexagentU_i \succ \edgeagentU_{e}^{\vertexagentU_i}[\Indp],\\
    \hat{\edgeagentU}_{e}^{\vertexagentU_i}[\Indp]&\colon \hat{\edgeagentW}_{e}^{\vertexagentU_i}[\Indp] \succ \edgeagentW_{e}^{\vertexagentU_i}[\Indp],
    &
    \hat{\edgeagentW}_{e}^{\vertexagentU_i}[\Indp] &\colon \edgeagentU_{e}^{\vertexagentU_i}[\Indp]  \succ \hat{\edgeagentU}_{e}^{\vertexagentU_i}[\Indp],\\[1ex]
    \hat{\edgeagentU}_{e}^{\vertexagentU_j}[\Indp]&\colon \hat{\edgeagentW}_{e}^{\vertexagentU_j}[\Indp]  \succ \edgeagentW_{e}^{\vertexagentU_j}[\Indp],
    &
    \hat{\edgeagentW}_{e}^{\vertexagentU_j}[\Indp] &\colon \edgeagentU_{e}^{\vertexagentU_j}[\Indp] \succ \hat{\edgeagentU}_{e}^{\vertexagentU_j}[\Indp],\\
    \edgeagentU_{e}^{\vertexagentU_j}[\Indp]&\colon \edgeagentW_{e}^{\vertexagentU_j}[\Indp] \succ \yW_e \succ \hat{\edgeagentW}_{e}^{\vertexagentU_j}[\Indp],
    &\qquad\qquad \edgeagentW_{e}^{\vertexagentU_j}[\Indp] &\colon \hat
    {\edgeagentU}_{e}^{\vertexagentU_j}[\Indp] \succ \vertexagentU_j \succ \edgeagentU_{e}^{\vertexagentU_j}[\Indp].
  \end{alignat*}
  To complete the construction, let $\Msize=|U|$, meaning that we are aiming for a perfect matching with minimum number of blocking pairs.
  Let $\Bnr=\kIS + \binom{\kIS}{2}$.
  The statement follows from the following claims.
  \begin{claim}\label{claim:IS->SMIbp}
    If $G$ admits a clique of size~$\kIS$, then there exists a perfect matching with at most $\Bnr$ blocking pairs.
  \end{claim}

  \begin{proof}\renewcommand{\qedsymbol}{(end of~\cref{claim:SMIbp->IS})~$\diamond$}
    Let $K$ be a size-$\kIS$ clique and let $E'$ be the induced edges in~$G[K]$.
    Without loss of generality, assume that $K=\{v_{1},\dots, v_{\kIS}\}$
    and $E'=\{e_{1}, \dots, e_{\binom{\kIS}{2}}\}$.
    We construct a matching~$\MM$ as follows.
    \begin{compactenum}[(1)]
      \item %
      Match the agents in $\{\hat{\vertexagentU}_{1}, \dots, \hat{\vertexagentU}_{k}\} \cup \{\vertexselectorW_1, \dots, \vertexselectorW_{\kIS}\}$ according to a perfect stable matching of $I$ restricted to $\{\hat{\vertexagentU}_{1}, \dots, \hat{\vertexagentU}_{k}\} \cup \{\vertexselectorW_1, \dots, \vertexselectorW_{\kIS}\}$; since the acceptability graph of the restricted to these agents is complete, we can find such a matching in polynomial time with Gale-Shapley algorithm~\cite{gale1962college}. 
Do the same for the set $\{\vertexselectorU_{1}, \dots, \vertexselectorU_{k}\} \cup \{\vertexagentW_1, \dots, \vertexagentW_k\}$.
For every $v_j \in K$, let $\MM(\vertexagentU_{j})=\hat{\vertexagentW}_{j}$.
      \item\label{match:nonCliquevertex} For each vertex~$v_i\notin K$, let $\MM(\hat{\vertexagentU}_{i})=\hat{\vertexagentW}_{i}$, and $\MM(\vertexagentU_{i})=\vertexagentW_{i}$.
      \item\label{match:Cliqueedge}
      Match the agents in $\{\edgeselectorW_{1}, \dots, \edgeselectorW_{\binom{\kIS}{2}}\} \cup \{\xU_1, \dots, \xU_{\binom{\kIS}{2}}\}$ according to a perfect stable matching of $I$ restricted to $\{\edgeselectorW_{1}, \dots, \edgeselectorW_{\binom{\kIS}{2}}\} \cup \{\xU_1, \dots, \xU_{\binom{\kIS}{2}}\}$; since the acceptability graph of the restricted to these agents is complete, we can find such a matching in polynomial time using the Gale-Shapley algorithm~\cite{gale1962college}.
      Do the same for the set $\{\edgeselectorU_{1}, \dots, \edgeselectorU_{\binom{\kIS}{2}}\} \cup \{\yW_1, \dots, \yW_{\binom{\kIS}{2}}\}$.
      For each edge~$e_{\Ind}\in E'$ and each incident vertex~$v_i$, and each $\Indp\in [L]$,
      let $\MM(\edgeagentU^{\vertexagentU_i}_{e_{\Ind}}[\Indp]) = \edgeagentW^{\vertexagentU_i}_{e_{\Ind}}[\Indp]$ and 
      $\MM(\hat{\edgeagentU}^{\vertexagentU_i}_{e_{\Ind}}[\Indp]) = \hat{\edgeagentW}^{\vertexagentU_i}_{e_{\Ind}}[\Indp]$.
      \item\label{match:noCliqueedge} For each edge~$e\notin E'$ and each incident vertex~$v_i$,  and each $\Indp\in [L]$, do the following.
      Let  $\MM(\xU_e) = \yW_e$,   
      $\MM(\edgeagentU^{\vertexagentU_{i}}_e[\Indp]) = \hat{\edgeagentW}^{\vertexagentU_{i}}_e[\Indp]$ and
      $\MM(\hat{\edgeagentU}^{\vertexagentU_{i}}_e[\Indp]) = \edgeagentW^{\vertexagentU_{i}}_e[\Indp]$.
    \end{compactenum}
    It remains to show that $\MM$ induces at most~$\Bnr$ blocking pairs.
    Let us consider every agent from~$U$.
    \begin{compactitem}[--]
      \item No vertex selector agent can be involved in a blocking pair since they only prefer a vertex agent with lower index, but $M$ restricted to the vertex selectors and vertex agents is stable by construction.
      \item Similarly, no edge selector agent can be involved in a blocking pair.
      \item Let $v_i\notin K$ be an arbitrary non-clique vertex. 
      Agent~$\hat{\vertexagentU}_i$ is already matched with his most preferred agent.
      Agent~$\vertexagentU_i$ prefers~$\hat{\vertexagentW}_i$ to his partner~${\vertexagentW}_i$, but agent $\hat{\vertexagentW}_i$ is matched with his most preferred agent; see \eqref{match:nonCliquevertex}.
      Agent~$\vertexagentU_i$ also prefers each agent~$\edgeagentW^{\vertexagentU_i}_e[\Indp]$ to his partner~${\vertexagentW}_i$.
      But each~$\edgeagentW^{\vertexagentU_i}_e[\Indp]$ is matched with his most preferred agent~$\hat{\edgeagentU}^{\vertexagentU_i}_e[\Indp]$ since no edge incident to~$v_i$ is in $E'$; see \eqref{match:noCliqueedge}.
      \item Let $e\notin E'$ be an arbitrary non-clique edge.
      Agent~$\xU_e$ is already matched with his most preferred agent~$\yW_e$.
      \item Let $e\notin E'$ be an arbitrary non-clique edge and let $\Indp\in [L], v_i \in e$.
      Agent~$\edgeagentU^{\vertexagentU_i}_e[\Indp]$ prefers~$\yW_e$ to his partner, but agent~$\yW_e$ is already matched with his most preferred partner~$\xU_e$; see \eqref{match:noCliqueedge}.
      Agent~$\edgeagentU^{\vertexagentU_i}_e[\Indp]$ also prefers~$\edgeagentW^{\vertexagentU_i}_e[\Indp]$ to his partner, but $\edgeagentW^{\vertexagentU_i}_e[\Indp]$ is already matched with his most preferred partner.
 Agent~$\hat\edgeagentU^{\vertexagentU_i}_e[\Indp]$ prefers~$\hat\edgeagentW^{\vertexagentU_i}_e[\Indp]$ to his partner, but $\hat\edgeagentW^{\vertexagentU_i}_e[\Indp]$ is already matched with his most preferred partner.
      \item Let~$v_i\in K$ be an arbitrary clique vertex.
      Agent~$\vertexagentU_i$ is matched to his most preferred agent.
      Agent~$\hat{\vertexagentU}_i$ forms with~$\hat{\vertexagentW}_i$ a blocking pair.
      He does not form with any vertex selector agent a blocking pair by reasoning analogously the first case.
      \item Let $e\in E'$ be an arbitrary clique edge.
      Agent~$\xU_e$ forms with $\yW_e$ a blocking pair.
      He does not form with any edge selector agent a blocking pair.
      For each incident vertex~$v_i$ of~$e$ (i.e., $v_i\in K$) and each $\Indp\in [L]$,
      both edge agents~$\edgeagentU_e^{\vertexagentU_i}[\Indp]$ and $\hat{\edgeagentU}_e^{\vertexagentU_i}[\Indp]$ are already matched with their most preferred agents, respectively.
      So they will not form with anyone a blocking pair.
    \end{compactitem}    
    Summarizing, all but the last two cases yield blocking pairs.
    The last two cases give~$\kIS$ and $\binom{\kIS}{2}$ blocking pairs, respectively.
    Hence, $\Bnr=\kIS+\binom{\kIS}{2}$, as desired.
  \end{proof}

  Before we continue with the backward direction, let us observe the following.
  \begin{claim}\label{claim:perfectmatching}
    Let $\MM$ be a perfect matching of the reduced instance.
    Then, for each edge~$e\in E$, it holds that $\MM(\yW_e)=\xU_e$ or $\MM(\yW_e)=\edgeselectorU_{\Ind}$ for some $\Ind\in [\binom{\kIS}{2}]$.
  \end{claim}

  \begin{proof}
    \renewcommand{\qedsymbol}{(end of~\cref{claim:perfectmatching})~$\diamond$}
    Since $\MM$ is perfect, by the preferences of the edge selectors, we know that there are exactly $\binom{\kIS}{2}$ edge-agents from~$U$ that are each matched with some edge selector~$\edgeselectorW_{\Ind}$ from~$W$.
    The remaining~$\emm-\binom{\kIS}{2}$ edge-agents~$\xU_e$ must be matched with their only available partner~$\yW_e$.
    This leaves~$\binom{\kIS}{2}$ edge-agents from~$W$ that are to be matched with some edge selector agents from~$U$, as desired.
  \end{proof}
  
  \begin{claim}\label{claim:SMIbp->IS}
    If $G$ does not admit a clique of size~$\kIS$, then every perfect matching induces at least $L> F(\Bnr)\cdot \Bnr$ blocking pairs. 
  \end{claim}

  \begin{proof}\renewcommand{\qedsymbol}{(end of~\cref{claim:SMIbp->IS})~$\diamond$}
    We show the contra-positive. 
    Let $\MM$ be a perfect matching of the reduced instance with at most~$L-1$ blocking pairs.
    Define~$K=\{v_i \mid \MM(\hat{\vertexagentU}_i)= \vertexselectorW_{\Ind} \text{ for some } \Ind \in [\kIS]\}$ and 
    $E'=\{e \mid \MM(\yW_e) \in \edgeselectorU_{\Ind} \text{ for some } \Ind \in [\binom{\kIS}{2}]\}$.
    We aim to show that $E'$ induces a size-$\kIS$ clique.
    
    Clearly $|K| = \kIS$ and $|E'|=\binom{\kIS}{2}$ since $\MM$ is a perfect matching.
    
    Consider an arbitrary edge~$e\in E'$ and let $e=\{v_i,v_j\}$.
    We claim that both~$v_i$ and $v_j$ must be in~$K$.
    
    Towards a contradiction, suppose that one of the endpoints of~$e$ is not in~$K$, say $v_i \notin K$.
    Then, by definition and by the preferences of $\vertexagentU_i$ we infer $\MM(\hat{\vertexagentU}_i) = \hat{\vertexagentW}_i$.
    By the acceptability graph over~$\vertexagentU_i$, we infer that either $\MM(\vertexagentU_i)=\vertexagentW_i$ or $\MM(\vertexagentU_i)\in \edgeagentWW(v_i)$.

    If $\MM(\vertexagentU_i)=\vertexagentW_i$,
    then for each~$\Ind\in [L]$, either $\{\vertexagentU_i, \edgeagentU^{\vertexagentU_i}_e[\Ind]\}$ or $\{\yW_e, \edgeagentW^{\vertexagentU_i}_e[\Ind]\}$ is blocking, inducing at least $L$ blocking pairs, a contradiction.

    If $\MM(\vertexagentU_i)=\edgeagentW_{e'}^{\vertexagentU_i}[\Indp]$ for some~$e'\in E$ with $v_i\in e'$ and $\Indp\in [L]$, then by the preferences of~$\hat{\edgeagentU}_{e'}^{\vertexagentU_i}$ we infer that $M(\hat{\edgeagentU_{e'}}^{\vertexagentU_i}) = \hat{\edgeagentW}^{\vertexagentU_i}_{e'}$.
    Since both $\edgeagentW^{\vertexagentU_i}_{e'}$ and $\hat{\edgeagentW}^{\vertexagentU_i}_{e'}$ are not available,
    we infer that $M(\hat{\edgeagentU}_{e'}^{\vertexagentU_i}) = \yW_{e'}$, a contradiction to~\cref{claim:perfectmatching}.

    We have just shown that for each edge~$e\in E'$ both endpoints must be in~$K$.
    Since $|E'|=\binom{\kIS}{2}$, we infer that $|K| \ge \kIS$.
    Since $|K|=\kIS$, we immediately obtain that $K$ induces a clique of size~$\kIS$.
  \end{proof}
  Since \Clique{} is \wone-hard, the theorem follows directly from \cref{claim:IS->SMIbp,claim:SMIbp->IS}.
\end{proof}

We proceed with the roommates setting.
We show that the same inapproximability result from the marriage setting holds for roommates setting even if we do not set any bound of the size of the target matching.
The idea is to adapt the reduction from \cref{thm:asm_inapprox} by adding a very large but \emph{even} number of agents for every original agent with preferences so that each original agent must be matched as desired in that reduction in order to avoid too many blocking pairs. 
Note that such adaption has be become a standard technique for many reduction; see for instance the work by \citet[Lemma 2]{abraham2005almost} and \citet[Construction 2]{chen_roommates}. 

\begin{corollary}\label{thm:asr_inapprox}
 Let $F\colon \mathds{N} \to \mathds{N}$ be an arbitrary computable function.
  Given an instance~$(\ppp, \Msize, \Bnr)$ of \minBPSRIdec\, it is \wone-hard to distinguish between the following two cases:
  \begin{compactenum}[(i)]
    \item $\ppp$ admits a $\Msize$ matching with at most~$\Bnr$ blocking pairs.
    \item Every matching of $\ppp$ has at least $F(\Bnr)\cdot \Bnr$ blocking pairs.
  \end{compactenum}%
\end{corollary}

\begin{proof}
  As mentioned above, we will modify the reduction from \cref{thm:asm_inapprox}. 
  Let $(G, \kIS)$ be an instance of \Clique, and let $(I,\Msize)$ be the instance of \minBPSMI constructed in \cref{thm:asm_inapprox}.
  Let $\hat \agents$ be the set of agents in $I$.
  Note that the reduction in \cref{thm:asm_inapprox} does not work for \minBPSRI, because \minBPSRI does not require every agent to be matched. %
  To fix this, we add a large set~$S_u$ of even number of agents for each original agent~$u$ in~$\hat{\agents}$ and append~$S_u$ to the end of the preferences of~$u$ to ensure that $u$ is matched in the same way as in the proof for~\cref{thm:asm_inapprox}.
  Let $\hat{I}$ be the modified instance for~\minBPSRI. 
  Then, 
  \begin{compactenum}[(i)]
\item if $M$ is a matching of $I$ in which every agent in $\hat \agents$ is matched to another agent in $\hat \agents$, then there is a matching $M'$ of $\hat I$ that admits the same number of blocking pairs as $M$, and
\item if $M'$ is a matching of $I'$ in which some agent in $\hat \agents$ is not matched to another agent in $\hat \agents$, then $M'$ admits more than $F(\kIS)$ blocking pairs.
\end{compactenum}
Thus, the corollary follows.
\end{proof}

\section{\fpt-approximability of \maxSMTI}

In this section, we complement the result by Marx and Schlotter~\cite{marx2010parameterized} and show that \maxSMTI admits an \fpt-\AS\ with respect to the number~$\tied$ of agents with preferences containing ties.
More precisely, we provide an \fpt-algorithm (with respect to~$\tied$) that for every approximation ratio~$\varepsilon\in [\frac{2}{3}, 1)$ outputs a stable matching with size at least~$\varepsilon\cdot \OPT(I)$ where $\OPT(I)$ denotes the size of an maximum-size stable matching in~$I$.
Note that we do not need to consider approximation factor between~$0$ and $\frac{2}{3}$ since there is a factor-$\frac{2}{3}$ polynomial-time algorithm by~\citet{KiralY2011} and \citet{mcdermid20093}.
The idea of our algorithm is to show that for each approximation ratio~$\varepsilon$,  either the size of a largest matching is upper-bounded by a function in~$(\tied, \varepsilon)$ or the approximated result by \citet{KiralY2011} and \citet{mcdermid20093} is good enough.
In the former case, we apply the kernelization algorithm by \citet{adil2018parameterized} to obtain an \fpt-algorithm for our parameter~$(\tied, \varepsilon)$.
\begin{theorem}
  \maxSMTI admits an \fpt-\AS\ with respect to the number~$\tied$ agents with ties. %
\end{theorem}

\begin{proof}
  Let $I$ be an instance of \maxSMTI, $\tied$ be the number of agents with ties, and $\varepsilon \in [0, 1)$ the approximation ratio we desire.
  As mentioned before, by  the results of~\citet{KiralY2011} and \citet{mcdermid20093}, we can assume without loss of generality that $\varepsilon \ge \frac{2}{3}$.
  Further, let $\MM^*$ be a maximum-size matching of the acceptability graph for~$I$.  
  We first observe the following relation between every stable matching and the size of~$\MM^*$.
  \begin{claim}\label{cla:tied_matching_size}
    For every stable matching of $I$, we have that $|\MM^*| \leq |\MM| + \tied$.
  \end{claim}
  
  \begin{proof}\renewcommand{\qedsymbol}{(end of \cref{cla:tied_matching_size})~$\diamond$}
    Consider the symmetric difference~$\MM^* \Delta \MM$.
    Each connected component in the symmetric difference is either a path and a cycle.
    Every cycle and even-length path component matches the same number of agents in both~$\MM$ and~$\MM^*$.
    Thus every component where~$\MM^*$ matches more agents must be an odd-length path.
    Moreover, each of these components matches precisely one extra pair.
    
    Assume, towards a contradiction, that there are more odd-length paths than agents with ties.
    Then there must be some odd-length path component where no agent has a tie in his preference list.
    
    Let $(v_1, \dots, v_t)$ be such a path, where $t$ is even.
    In $\MM^*$, the pairs $\{a_1, a_2\}, \dots,$ $\{a_{t - 1}, a_t\}$ are matched, whereas in $\MM$, the pairs $\{a_2, a_3\}, \dots, \{a_{t - 2}, a_{t - 1}\}$ are matched, and $a_1, a_t$ are unmatched.
    Since $\MM$ is stable, it cannot be blocked by $\{a_1, a_2\}$. As $a_1$ is unmatched in $\MM$, it must be that $a_2 \colon a_3 \succ a_1$. Since $\MM^*$ is stable, it cannot be blocked by $\{a_2, a_3\}$, and we must have that $a_3 \colon a_4 \succ a_2$. By repeating this argument, we obtain that $a_{t - 1} \colon a_t \succ a_{t - 2}$. But then $\{a_{t - 1}, a_t\}$ blocks~$\MM$, a contradiction.
  \end{proof}

  Let $\MM$ be a stable matching with $|\MM| \geq \frac{2}{3}|\MM^*|$.
  Note that we can find $\MM$ in polynomial time by using for instance the algorithm of \citet{mcdermid20093}.
  We proceed in two cases:
  
  \begin{description}
    \item[Case 1:]$|M| <  \frac{\varepsilon\cdot \tied}{1-\varepsilon}$.
    Then, 
    $|\MM^*| \leq \frac{3}{2}|\MM| < \frac{3\varepsilon\cdot \tied}{2(1-\varepsilon)}$.
    Since \maxSMTI has a size-$O(\Msize^2)$ problem kernel~\cite{adil2018parameterized} where $\Msize$ is the target size of a stable matching in the natural parameterization of \maxSMTI,
    and $\Msize$ is bounded by $|\MM^*|$,
    we can solve $I$ exactly in \fpt-time with respect to $\tied + \varepsilon$.
    \citet{adil2018parameterized} also provide an \fpt-algorithm with running time~$\Msize^{O(\Msize)}+n^{O(1)}$.
    Applying it, we obtain our \fpt-algorithm with running time~$\frac{3\varepsilon\cdot \tied}{2(1-\varepsilon)}^{O(\frac{3\varepsilon\cdot \tied}{2(1-\varepsilon)})}
    +|I|^{O(1)}$.
    \item[Case 2:]
    $|\MM| \geq  \frac{\varepsilon\cdot \tied}{1-\varepsilon}$.
    Combining with \cref{cla:tied_matching_size},
    $|M^*| \le |M| + \frac{(1-\varepsilon)}{\varepsilon} \cdot |M| = \frac{|M|}{\varepsilon}$. 
    Thus, $|M|$ is an $\varepsilon$-approximation.
  \end{description}
  
  As both cases lead to a $\varepsilon$-approximation in \fpt-time, this concludes the proof.
\end{proof}

\section{Conclusion}
In this manuscript, we have explored the fixed-parameter approximability of three problems in the field of stable matchings.
Our goal is to highlight the potential of fixed-parameter approximability in the field of matching under preferences and computational social choice more generally.
This technique can be used to solve problems that resist both polynomial-time approximation and \fpt-algorithms.
Using this technique more widely in the field of matchings and computational social choice is an open question. 
Of course, as we have in \cref{thm:asm_inapprox,thm:asr_inapprox}, some problems resist even this.
Finding suitable parameters is part of the challenge.

\bibliographystyle{plainnat}
\bibliography{bib}

\end{document}